\documentclass{article}
\usepackage[utf8]{inputenc}
\usepackage{hyperref}

\usepackage{lipsum,graphicx,float}
\usepackage{subfig,caption}
\usepackage[margin=3cm]{geometry}
\usepackage{amsthm}
\usepackage{amssymb}
\usepackage{placeins}

\floatstyle{ruled}
\newfloat{algorithm}{tbp}{loa}
\providecommand{\algorithmname}{Algorithm}
\floatname{algorithm}{\protect\algorithmname}
\newtheorem{thm}{\protect\theoremname}
\newtheorem{example}[thm]{\protect\examplename}
\usepackage{listings}
\providecommand{\examplename}{Example}
\providecommand{\theoremname}{Theorem}

\title{Towards a global dynamic wind atlas: A multi-country validation of wind power simulation from MERRA-2 and ERA-5 reanalyses bias-corrected with the Global Wind Atlas}
\author{Katharina Gruber, Peter Regner, Sebastian Wehrle, Marianne Zeyringer, Johannes Schmidt}
\date{December 2020}

\usepackage[backend=biber, style=ieee, citestyle=numeric-comp, url=false, doi=false, isbn=false, dashed=false]{biblatex}
\usepackage{graphicx}

\begin{document}

\maketitle

\begin{abstract}
Reanalysis data are widely used for simulating renewable energy and in particular wind power generation. While MERRA-2 has been a de-facto standard in many studies, the newer ERA5- reanalysis recently gained importance. Here, we use these two datasets to simulate wind power generation and evaluate the respective quality in terms of correlations and errors when validated against historical wind power generation. However, due to their coarse spatial resolution, reanalyses fail to adequately represent local climatic conditions. We therefore additionally apply mean bias correction with two versions of the Global Wind Atlas (GWA) and assess the respective quality of resulting simulations. Potential users of the dataset can also benefit from our analysis of the impact of spatial and temporal aggregation on simulation quality indicators. While similar studies have been conducted, they mainly cover limited areas in Europe. In contrast, we look into regions, which globally differ significantly in terms of the prevailing climate: the US, Brazil, South-Africa, and New Zealand. Our principal findings are that (i) ERA5 outperforms MERRA-2, (ii) no major improvements can be expected by using bias-correction with GWA2, while GWA3 even reduces simulation quality, and (iii) temporal aggregation increases correlations and reduces errors, while spatial aggregation does so only consistently when comparing very low and very high aggregation levels.

\end{abstract}

\section{Introduction}
Decarbonising global energy systems is at the core of climate change mitigation. The expansion of renewable energies is one important measure to attain this goal \cite{Jaegemann2013,Dagoumas2019}.

Globally, wind power and solar PV have been the renewable energy sources with the highest growth rates in recent years. While the installed capacity on a global level is similar for PV (579 GW) and wind power (594 GW), wind power generation (1195 TWh) is substantially higher than electricity generation from PV (550 TWh) \cite{irena2020}. 
This trend of a higher share of wind power generation is likely to continue for some world regions, e.g. Europe \cite{ECA2019}. Scenarios have explored the importance of wind power in future energy systems, with shares of around 50\% of global power demand in 2030 \cite{Jacobson2011}, 74\% in Europe by 2050 \cite{Zappa2018}, or even 80\% to 90\% of the European VRES mix \cite{Eriksen2017}.\\
For an adequate assessment of the impacts of high shares of renewable electricity and in particular of wind power generation on power systems, long spatially and temporally highly resolved renewable power generation time series are necessary to represent short and long term changes in resource availability \cite{Collins2018}. At least one climatological normal of 30 years should be used to understand variability \cite{WMO2017}.\\

Reanalysis climate data sets are frequently used to generate such time series.
Two of the most prominent global reanalyses are NASA's MERRA and MERRA-2 and the more recent ERA5 provided by the European Centre for Medium-Range Weather Forecasts.
The MERRA reanalyses were used for example for estimating the global technical onshore and offshore wind power generation potentials \cite{Bosch2017,Bosch2018}, or the integration of renewables into the European power system \cite{Huber2014}. Also correlations between wind power generation in European countries \cite{Olauson_2016}, extreme events in Britain \cite{Cannon_2015}, or the impacts of uncertainty factors \cite{Monforti_2017} and ageing \cite{Soares2020} in wind power simulation were studied.
With ERA5 global \cite{Soares2020} and Lebanese \cite{IbarraBerastegi2019} offshore wind power potential, as well as electricity demand and renewable generation in Europe \cite{Bloomfield2020a} and West Africa \cite{Sterl_2018} were estimated.
While global reanalysis data sets offer the advantage of conducting multi-country or global analyses without the need for country or region-specific climate data sources, they also come with their drawbacks.
Although the temporal resolution is usually high at one hour or even less, the spatial resolution is rather coarse at a grid size of several kilometres (eg. MERRA-2 about 50 km). Therefore, those data sets, in contrast to regional reanalyses such as COSMO-REA \cite{CosmoREA2}, are limited in representing local climatic conditions in sufficient detail, as required for the simulation of wind power generation \cite{Staffell_2016}.
It is known that reanalysis data are subject to bias \cite{Cannon_2015,Pfenninger_2016,Olauson_2016}. To increase simulation quality, efforts should be made to correct the bias \cite{Monforti_2017,Henckes2020}, as the bias of reanalysis data may result in differences in model-derived installed capacities of up to 20\% difference \cite{Henckes2020}.
In many cases, however, reanalysis data is used directly \cite{Ren2019, Monforti_2017, Cannon_2015, Cradden2017, Kubik2013, Camargo2019, Camargo2019a}. If it is corrected, observed wind power generation data is mostly used \cite{Olauson_2018, Staffell_2016, Olauson_2015, Olauson_2016, Camargo2019b}. This approach is not globally applicable, as observations of wind power generation are unavailable for many world regions. Additionally, data quality and the level of temporal and spatial aggregation varies between countries.\\
Therefore, other forms of bias correction are required when conducting global analysis \cite{Staffell_2016}. Here, we aim at reducing the bias in reanalysis data by applying the Global Wind Atlas \cite{GWA3}. Recently, the Global Wind Atlas Version 3.0 has been released and we put a particular focus on assessing the quality of this latest version compared to the previous version 2.1. GWA 3.0 has - at the moment- only been assessed for Pakistan, Papua New Guinea, Vietnam, and Zambia for wind speeds \cite{GWA3val}, however not for the purpose of wind power simulation.\\

Of course, the GWA may not necessarily decrease bias. It is therefore of great interest to validate simulated wind power simulation data against observed generation - for both, raw reanalysis data and reanalysis data corrected with the GWA. Other work has mainly focused on validating raw wind power simulation data:
\citeauthor{Staffell_2016} validate wind power simulations derived from MERRA and MERRA-2 against observed generation data for 23 European countries and find significant bias. 
\citeauthor{Olauson_2015} \cite{Olauson_2015} used the MERRA data set to model Swedish wind power generation, and production data from the Swedish TSO to validate and bias-correct their modelled data. In a comparison of  MERRA-2 and ERA5 for the use of wind power simulation, time series for four European countries and one region in the USA were validated\cite{Olauson_2018}.
\citeauthor{Jourdier2020} compared MERRA-2 and ERA5 \cite{Jourdier2020} to simulations of French wind power generation based on two high-resolution models (COSMO-REA6 and AROME) and a mesoscale model (NEWA) and validated all datasets against observed wind speed and power generation data.\\
Since most of the previous analyses only assessed one particular reanalysis data set, we focus on the comparison of ERA5 and MERRA-2, on results quality and the additional use of the GWA for bias-correction.
As Europe has already been studied in several other analyses \cite{Staffell_2016,Olauson_2015, Jourdier2020,Monforti_2017,GonzalezAparicio2017} and to cover different global climatic conditions, we study the following non-European countries: Brazil, USA, South Africa and New Zealand. These countries are spatially very diverse, host significant wind power capacities, and provide timeseries of wind power generation that can be used for validation.
Furthermore, we contribute to a better understanding of the role of spatial and temporal resolution by assessing simulation quality on different levels of spatial and temporal aggregation. This is highly relevant information for users in power- and energy system models \cite{Bloomfield_2020}. 

In particular, we answer the following research questions: (1) Does the newer reanalysis ERA5 with higher spatial resolution perform better than the older MERRA-2 when validated against historical wind power generation data? (2) Does bias-correction with the spatially highly resolved GWA increase simulation quality? (3) Does the GWA 3.0 perform better than the previous GWA 2.1.? (4) Does aggregating single wind parks to larger systems decrease the error due to spatial complementarity and error compensation effects, as indicated by Goi\'{c} et al. \cite{Goic2010} and Santos-Alamillos et al. \cite{SantosAlamillos2015}? (5) Does temporal aggregation reduce errors?
We assess those questions by simulating wind power generation in the four countries for all wind parks, using both ERA5 and MERRA-2 with and without bias-correction with the GWA. We validate simulated against observed generation on different spatial levels and compare quality between all simulations.

\section{Data}
We use several data sets for simulation, bias correction and validation: wind speeds are taken from the MERRA-2 and ERA5 reanalysis data sets. The GWA is used for mean bias correction. Information on wind park locations and the used turbine technology is collected from different country specific data sources (see section \ref{subsection:windpark_info}). Similarly, country specific wind power generation data is gathered to perform the final validation. \\

\subsection{Reanalysis data}
From MERRA-2 \cite{MERRA2}, we use the time-averaged, single-level, assimilation, single-level diagnostics (tavg1\_2d\_slv\_Nx) dataset, while we use hourly data on single levels from 1950 to present from ERA5\cite{ERA5}. MERRA-2 reanalysis data are provided by the National Aeronautics and Space Administration via the Goddard Earth Sciences Data and Information Services Center and follow the earlier version of MERRA, while ERA5 is the follow-up product of ERA-Interim provided by the European Centre for Medium Range Weather Forecast (ECMWF). MERRA-2 is available for circa 40 years (since 1980), while ERA5 has recently been extended to reach back to 1950. While both exhibit a temporal resolution of one hour, the spatial resolution is higher in the more recent ERA5 data set (~31 km) than in MERRA-2 (~50 km).\\
The climate input data is downloaded for time periods corresponding to the temporal availability of validation data. Spatial boundaries are defined by the size of the respective country. 
The downloaded parameters are eastward (u) and northward (v) wind speeds at two different heights for each reanalysis data set (ERA5: 10 m and 100 m above surface, MERRA-2: 10 m above displacement height and 50 m above surface), as well as the displacement height for MERRA-2.

\subsection{Global Wind Atlas}
The Global Wind Atlas \cite{GWA3} provided by the Technical University of Denmark (DTU) is used to spatially downscale the reanalysis data to a resolution of 250 m, in order to take into account local variations of mean wind speeds. The current version, GWA 3.0 was derived from the ERA5 reanalysis and provides mean wind speeds and mean power densities at five different heights (10, 50, 100, 150 and 200 m), as well as mean capacity factors for three different turbine classes according to IEC\footnote{
International Electrotechnical Commission} for the period 2008-2017. Furthermore, there are layers describing the terrain surface and a validation layer showing in which countries and for which wind measurement stations the GWA has been validated.\\
The previous version, GWA 2.1, which is also used in this analysis, provides wind speeds at only three heights (50, 100 and 200 m) at the same spatial resolution and was derived from ERA-Interim, the preceding data set of ERA5 \cite{Badger2019} for the period 1987-2016.\\
For the purpose of mean bias correction, the wind speed layers at 50 m and 100 m height are downloaded for each country. They correspond to the upper layer of reanalysis wind speeds in MERRA-2 and ERA5, respectively. Since the GWA2 is no longer available at the official GWA homepage, data were extracted from the stored global data set \cite{GWA2} around the country boundaries.

\subsection{Wind park information}
\label{subsection:windpark_info}
\begin{figure}[!ht]
\centering
\includegraphics[scale=0.7]{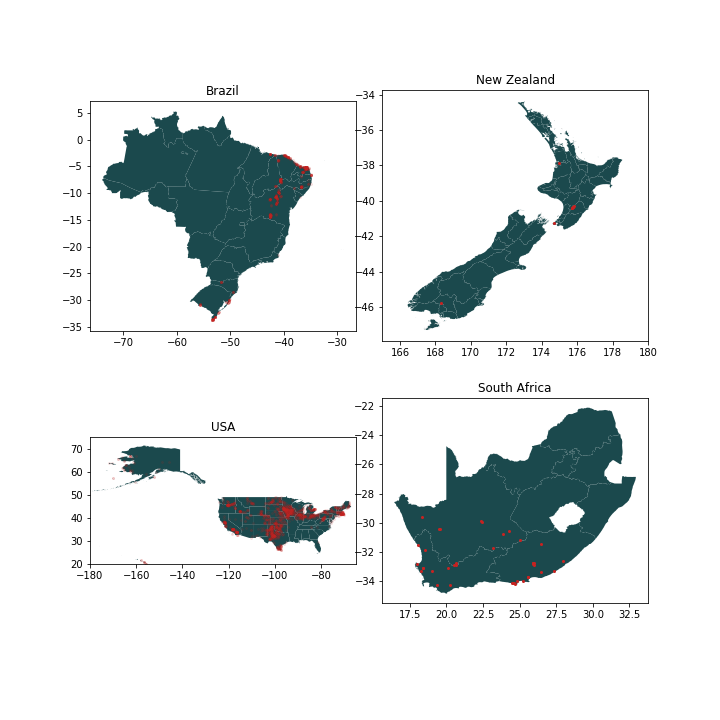}
\caption{Locations of wind parks in Brazil, New Zealand, USA and South Africa}
\label{fig:wp_map}
\end{figure}

For the simulation of wind power generation, we use turbine specific information on location, installed capacity, hub height and rotor diameter. The spatial distribution of wind power plants is shown in Figure \ref{fig:wp_map}. In countries where turbine specific location information is not available, we use wind park specific data. This information is retrieved from freely available country level data sets (see Table \ref{tab:TURB_table}).\\
For Brazil, two data sets, the Geographic Information System of the Electrical Sector (SIGEL) \cite{SIGEL} and the Generation Database (BIG) \cite{BIG}, from the National Electrical Energy Agency (ANEEL) \cite{ANEEL} are combined using the wind park codes.
The use of both datasets is necessary, as SIGEL data contains only the location, installed capacity, hub height and rotor diameter, while the state and the commissioning dates are added from the BIG database.
Two wind turbines in the BIG dataset have a hub height and rotor diameter of 0 meters. They are replaced by values from turbines with similar capacity.\\
The information on ten wind parks with available production data is collected from the New Zealand Wind Energy Association \cite{NZWEA}. Similarly, the information on 39 wind parks in South Africa is gathered from the Renewable Energy Data and Information Service (REDIS) \cite{ZAFwp}, while rotor diameters, hub heights and capacities are complemented with information from The Wind Power\cite{TWP}. Since several data points were obviously erroneous or missing, the database was completed with an online search (see Table \ref{tab:zaf_turb_complete}). The resulting South Africa wind park data set is available online for further use \cite{GDWA}.\\
The information on the over 60 000 wind turbines in the USA is obtained from the US Wind Turbine Data Base (Version 3.2) \cite{USWTDB}, which comprises most of the necessary data. Missing information\footnote{Lacking data of commissioning date: 1540 turbines, turbine capacity: 5530 turbines, hub height: 7790 turbines, and rotor diameter: 6728 turbines} is replaced by the yearly mean (installed capacities, hub heights) or the overall mean (commissioning year) and rotor diameters are completed by fitting a linear model to the hub heights. In some cases the specific power calculated from rotor diameter and capacity is too low (below 100 W/m\textsuperscript{2}) resulting in unrealistic power curves, and thus replaced by the mean specific power of turbines with the same capacity \footnote{This applies to 49 wind turbines, of which 48 have incomplete turbine specifications}.

\begin{table}[ht]
    \centering
    \caption{Wind turbine and wind park data sets applied for simulation}
    \makebox[\textwidth][c]{}
    \begin{tabular}{p{1.3cm}p{1.3cm}p{1.3cm}p{1.3cm}p{1.3cm}p{1.3cm}p{1.3cm}p{1.3cm}p{1.3cm}p{1.3cm}}
    \hline
    Country      & Source                                       & Avail-ability  
        & turbines  & parks & total capacity [MW]   & avg. park capacity [MW]
        & avg. turbine capacity [kW]    & avg. rotor diameter [m]   & avg. hub height [m] \\ 
    \hline
    \hline
    Brazil       & ANEEL (BIG, SIGEL) \cite{ANEEL,BIG,SIGEL}    & turbines      
        & 7438      & 603   & 15190                 & 25
        & 2031                           & 98                        & 87 \\
    New Zealand  & NZWEA \cite{NZWEA}                           & wind parks    
        & 405       & 10     & 564                   & 56
        & 1719                           & 61                        & 53 \\
    South Africa & REDIS \cite{ZAFwp} and various               & wind parks    
        & 1466      & 39   & 3545                 & 90
        & 1719                           & 84                        & 95 \\
    USA          & USWTDB \cite{USWTDB}                         & turbines      
        & 63002     & 1565 & 108301                 & 69
        & 2525                          & 105                       & 75 \\
    \hline
    \end{tabular}
    \label{tab:TURB_table}
\end{table}

\subsection{Wind power generation data for validation}
The validation of the simulated wind power generation time series is based on observed generation at different spatial and temporal resolutions, gathered from country specific data sources. While there is data available on all time scales (hourly, daily and monthly) for each of the four studied countries or regions in those countries, historical wind power generation records on the level of wind parks are available only for Brazil and New Zealand. In South Africa, the country's observed power generation is only available per Cape (Eastern, Northern and Southern Cape), while for the USA the smallest level of spatial disaggregation available is the state level.\\
Temporal availability of the generation time series varies depending on the data source and commissioning dates of wind parks. The highest resolution of data  is given in Brazil, where the National Electrical System Operator (ONS) \cite{ONS} provides data on three temporal (hourly, daily, monthly), as well as four spatial levels (wind park, state, subsystem, country). Of the 174 wind parks in Brazil for which hourly data are available in the ONS dataset, 70 can be matched by their name to simulated wind parks based on ANEEL data, and 42 show sufficient data quality (also see Table \ref{tab:data_cleaning_bra}). They are consequently used for the further analysis. Due to data quality issues and the requirement of consistency only hourly data on the wind park level were used and aggregated spatially and temporally (also see section \ref{subsection:data_cleaning}).
In New Zealand, wind park specific generation data is also available, however only for ten wind parks. The information on historical wind power and other generation is provided by the Electricity Market Information (EMI) \cite{EMI} half hourly and aggregated to hourly production values for validation against hourly simulated values.\\
In South Africa, generation data is provided by REDIS \cite{REDIS} as capacity factors. For observed power generation in the USA, several data sources are used. The U.S. Energy Information Administration (EIA) \cite{EIA} provides monthly resolved generation data for the USA, its 51 states and 10 sub-regions\footnote{New England, Mid-Atlantic, East North Central, West North Central, South Atlantic, East South Central, West South Central, Mountain, Pacific Continental and Pacific Non-Continental}. For New England\footnote {Connecticut, New Hampshire, Maine, Massachusetts, Rhode Island and Vermont}, monthly data are retrieved from ISO New England \cite{isoNE}\footnote {Data from EIA were discarded due to poor quality (nearly constant/fluctuating generation instead of seasonal pattern and some very low production months, see Figure \ref{fig:regionsm}) and instead ISO New England data are used}. The Electric Reliability Council of Texas (ERCOT) \cite{ERCOT} provides hourly generation data for Texas. The 5-minute wind power generation data in the Bonneville Power Administration (BPA) \cite{BPA}, which is responsible for 49 wind parks in the regions of Oregon and Washington, is aggregated to hourly output.\\
Table \ref{tab:valdata_tab} summarises the data sources used for validation.\\

\begin{table}[ht]
    \centering
    \caption{Data sets applied for validation}
    \begin{tabular}{llll}
    \hline
    Country      & Regions                          & Temporal resolution       & Source \\ 
    \hline
    \hline
    Brazil       & 42 wind parks, 4 states, country & hourly, daily, monthly    & ONS \cite{ONS}\\ \hline
    New Zealand  & 10 wind parks, country           & hourly, daily, monthly    & EMI \cite{EMI}\\ \hline
    South Africa & 3 capes, country                 & hourly, daily, monthly    & REDIS \cite{REDIS}\\ \hline
    USA          & 25 states, 8 regions, country & monthly                   & EIA \cite{EIA}     \\ 
                 & Texas                            & hourly, daily, monthly    & ERCOT \cite{ERCOT} \\
                 & New England                      & monthly                   & ISO New England \cite{isoNE} \\
                 & BPA                              & hourly, daily, monthly    & BPA \cite{BPA} \\
    \hline
    \end{tabular}
    \label{tab:valdata_tab}
\end{table}

\subsection{Data cleaning} 
\label{subsection:data_cleaning}
In a preliminary screening, parts of the available observed wind power generation time series showed long sequences of missing data and unlikely generation patterns, such as long periods of constant output. We therefore applied a thorough cleaning procedure.
\subsubsection{Brazil}
First, wind park names in the ANEEL and the ONS data set have to be machted in order to validate the simulation with observed generation from the according wind park. Due to the large number of available wind park data, this step is performed using fuzzy matching, ignoring special characters and case sensitivity. Only wind parks with a matching score of 100 are used for validation. From a total of 174 parks, only 72 satisfied this criterion. \\
For these wind parks, leading and trailing series of zero production are removed from hourly generation time series at wind park level. For constant parts of time series, two different approaches are taken. If those parts are 0, they either indicate (a) a long period of very low or very high wind speeds (i.e. either below cut-in or above cut-out wind speed), (b) a downtime of the turbine due to e.g. maintenance, and (c) an error in the observed data. Filtering out all instances of 0 wind power production would remove all three events, however, this would be inconsistent with other countries, where this approach cannot be taken (as wind power generation on the level of wind parks is not available). We therefore opted for removing constant parts of the timeseries with periods of 0 generation larger than the largest period of 0 generation in the simulated timeseries which accounts to 180 hours.\\
For other constant parts of the timeseries, which are above 0, we removed them if the period was longer than 24 hours. Time series which contain less than 2 years of data are excluded from the analysis to guarantee capturing seasonal effects. We stress that the two years of data do not necessarily occur consecutively.
Furthermore, the data are assessed with respect to their capacity factors. We removed all instances in the timeseries where capacity factors above 1 where observed. Table \ref{tab:data_cleaning_bra} gives an overview how many locations where affected by the performed data cleaning in Brazil.\\

\begin{table}[ht]
    \centering
    \caption{Data cleaning steps and remaining wind parks for validation in Brazil}
    \begin{tabular}{lrr}
    \hline
                & Applies to    & Remaining \\&&wind parks \\ 
    \hline
    \hline
    - total number of observed wind park time series 
                &               & 174 \\ \hline
    1. matching of ONS and ANEEL  
                &               & 72 \\ 
    - keep only 100 matching score	
                &               & 70 \\ \hline
    2. data cleaning & &\\
    - remove constant parts of time series except 0 ($>$24h) 
                & 50            & 70 \\ 
    - remove constant parts of 0 generation ($>$180h)
                & 28            & 70\\
    - remove capacity factors $>$ 1
                & 59            & 70\\
    - remove short time series ($<$2y)
                & 17            & 53 \\ 
    \hline
    \end{tabular}
    \label{tab:data_cleaning_bra}
\end{table}

In order to ensure consistent data quality throughout the evaluation, instead of applying the temporally and spatially aggregated data sets provided by ONS, we aggregate the hourly wind power generation time series on wind park level spatially and temporally. This is necessary since daily data are equal to aggregated hourly data on the ONS site. However, this approach ignores missing or erroneous data, resulting in lower power generation in periods where generation data are missing in at least one of the wind parks in a particular region. We remove time steps from simulation data, when the data are missing in some wind parks in the validation data and aggregate after this correction.
Furthermore, hourly and daily data are not consistent with monthly data. As the applied aggregation method is not made explicit, the reason for the inconsistency remains unclear. To overcome the inconsistency, aggregation of validation data is performed starting at the highest spatio-temporal resolution of the available data, i.e. at the hourly wind park data. This approach allows to remove missing data from all spatial and temporal scales, improving the fit of observed and simulated data.\\

\subsubsection{USA}
In the USA, different measures were applied depending on the data source. In the EIA data set, leading zero production is removed. Since before 2010 the fit of simulation to validation data is low, the installed capacity in the USA from the USWTDB is compared to the yearly cumulative installed wind power capacity as provided by IRENA \cite{IRENA}. This comparison shows large inconsistencies (see Figure \ref{fig:uswtdb_irena}). Therefore, wind power generation is analysed for the past ten years only, starting in 2010. This approach notably improves  results (see Figure \ref{fig:2000vs2010}).
Despite the cleaning measures, several regions still result in unusually low correlations and high errors. A visual inspection of the monthly time series shows that the observed generation of several states and regions is nearly constant or repetitively fluctuating between different generation levels for long time series. This contrasts with our expectation of observing seasonal patterns (see section \ref{subsection:quality_USA}). Due to this reason, seven states and three regions affected by this approach are discarded for further use, while in nine states, only part of the time series is used for validation. These are indicated in Figure \ref{fig:statesm}.
In the BPA data set, some observations are missing. As the data is available at a 5 minutes resolution, the missing values are interpolated. The maximum consecutive missing observations is one hour.\\

\subsubsection{New Zealand and South Africa}
In New Zealand, constant output over more than 24 hours is removed from the time series. No further data cleaning operations are applied. In South Africa, a limited amount of capacity factors larger than 1 are observed. These time steps are removed.\\

\section{Methods}

\subsection{Wind power simulation}
Wind power is simulated based on reanalysis data and mean wind speeds in the GWA. In a preparatory step, effective wind speeds are calculated from eastward (u) and northward (v) wind speed components in reanalysis data according to the Pythagorean theorem for the two heights available.
From the effective wind speed, the Hellmann exponent $\alpha$, describing the structure of the surface, is calculated. Using the location information of wind turbines or wind parks, reanalysis and GWA wind speeds are interpolated to the nearest neighbour and extrapolated to the hub height using Hellmann's power law.\\
When bias correction is applied, mean wind speeds are retrieved from the GWA at the location closest to the wind park or turbine and divided by the average of the reanalysis wind speed time series at the specific locations at the same height, i.e. 50 m for MERRA-2 and 100 m for ERA5, as these are the heights closer to hub height.
This quotient is used as a bias correction factor to shift reanalysis wind speeds interpolated to hub height up or down according to the GWA.\\
In order to convert wind speeds to wind power, the power curve model introduced by Ryberg et al. \cite{Ryberg2019} is applied and scaled to the installed capacity of the turbines. 

This model estimates power curves empirically from the specific power, i.e. the installed capacity per rotor swept area, of wind turbines. It therefore does take into account differences in the power output according to specific power, but additional technology or turbine specific effects are not considered. We follow this approach, as otherwise we would have to manually research power curves for 283 different turbine models, and as additionally turbine models are not know for 865 cases.
Wind power generation is simulated for the whole country-specific time period, but generation is set to 0 for periods before the commissioning date of the respective wind park. 
If only the month of commissioning is known, we assume the middle of the month as commissioning date. For the USA, only the commissioning year is known. 
In order to avoid large increments of wind power generation on any particular date, the capacity installed within a year is linearly interpolated from the 1st of January to the end of the year.\\

\subsection{Validation}
218 different data sets of observed generation are suitable for validation. 10 data sets are on country scale, 58 on state or regional scale, and 150 on wind park scale. 62 of those have hourly resolution, 62 daily, and 94 monthly. Due to data quality issues, not all available time series could be used (see section \ref{subsection:data_cleaning}).
In order for results to be comparable between different levels of spatial and temporal aggregation, as well as countries, generation time series are normalised to capacity factors.\\
Validation of the simulated time series was performed using three statistical parameters to assess quality. Pearson correlation, RMSE (root mean square error) and MBE (mean biased error) were used, as suggested by Borsche et al. \cite{Borsche2015}.

The RMSE is an indicator that increases if (a) there is a significant difference in the level of simulated and observed timeseries, and (b) if there is a temporal mismatch between the two. As we use capacitiy factors which are comparable in scale between regions, the RMSE does not have to be normalized. To assess the different components of mismatch, i.e. temporal mismatch and mismatch in level of production, we additionally calculate the Pearson correlation which indicates if the temporal profile of simulated and observed generation are similar. To assess differences in levels including over- or underestimation, we determine the MBE.

\begin{figure}[!ht]
\centering
\includegraphics[scale=0.5]{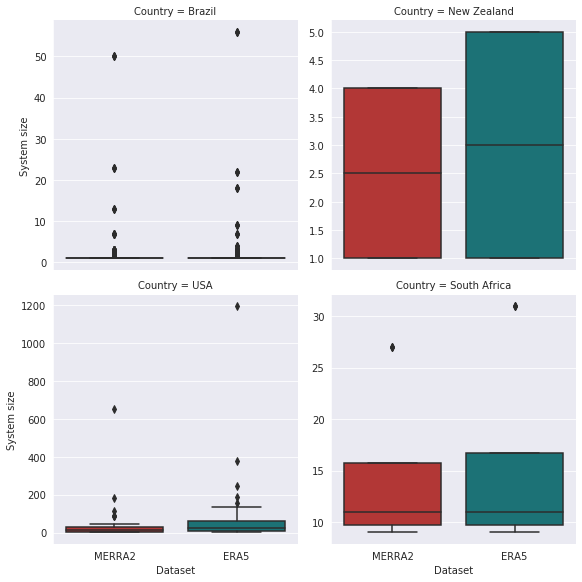}
\caption{System sizes per country and data set (non-normalised)}
\label{fig:syssize}
\end{figure}

Since the proposed model does not consider losses due to wakes or down-times due to maintenance, a slight overestimation of generation is expected. I.e. slightly overestimating models tend to represent actual generation better than underestimating ones. 
Results for different regions and temporal aggregation levels are compared in notched boxplots. The notches indicate if the median's differ significantly at the 95\% level \footnote{The notches are determined according to $M \pm 1.57 \cdot IQR \cdot \sqrt{n}$, with M being the median, IQR the interquartile range and n the number of samples. If the notches of two boxes do not overlap, the difference between their medians is statistically significant at the 0.05 level \cite{Chambers1983}.} As we cannot assume that our sample of wind parks and regions represents a random sample of global wind power generation locations and as there is a bias in the amount of timeseries available for different regions, we report on different results for different countries whenever they deviate from the generally observed pattern. Respective figures are put into the appendix.\\
In order to estimate the effect of system size on  simulation quality, a system size parameter is introduced.
It measures the number of reanalysis grid cells occupied by wind turbines or parks, e.g. per wind park or region (see Figure \ref{fig:syssize}). Individual wind turbines therefore always have size 1. Wind parks can have a size larger 1, if they cover more than one grid cell, but this is mostly not the case. Countries cover always more than one grid cell.\\

\section{Results}
\label{section:results}
In this section we first present how the choice of the reanalysis dataset affects simulation quality. Subsequently, we investigate whether the use of the GWA for mean bias correction can improve our simulation's goodness of fit. Finally, we assess the effect of spatial and temporal aggregation of wind power generation on simulation quality.

\subsection{Impact of choice of reanalysis dataset on simulation quality}


Here we assess the difference in simulation quality as implied by using different reanalysis data sets, i.e. MERRA-2 and the more recent ERA5.\\

\begin{figure}[!h]
\centering
\includegraphics[scale=0.7]{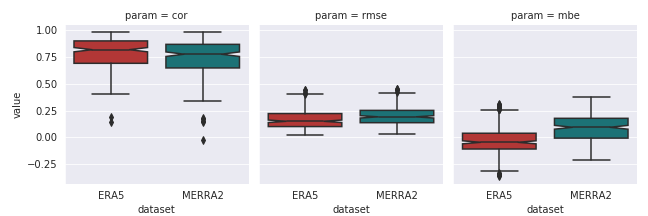}
\caption{Comparison of statistical parameters for simulations with ERA5 and MERRA-2 reanalyses for all analysed regions. Non-overlapping notches indicate a  difference in the medians statistically significant at the 95\% level.}
\label{fig:era5_vs_merra2_all}
\end{figure}

Figure \ref{fig:era5_vs_merra2_all} presents a comparison of statistical parameters between simulations based on ERA5 and MERRA-2 reanalyses for all analysed regions, i.e. wind parks, states, regions, and countries. While ERA5 correlations (median: 0.82) are higher than the ones achieved with MERRA-2 (median: 0.77) and while MERRA-2 has a larger spread of correlations, one of them being even negative, the difference in correlations is not significant. Overall, there is a significant (notches do not overlap) difference in RMSEs (median ERA5: 0.15, MERRA-2: 0.19). Regarding the MBEs, there is a significant difference between the median MBE of ERA5 (-0.05) and MERRA-2 (0.09), with ERA5 MBEs slightly underestimating generation on average, while MERRA-2 overestimating generation quite substantially (by approx. 1\%). Underestimation of ERA5 can be as low as almost 40\% for some locations, while MERRA2 overestimates generation by as much as 40\%. In general, both data sets seem to underestimate wind power generation in New Zealand, which is the only region where this occurs.\\
On a country level (see Figure \ref{fig:era5_vs_merra2_dif}), these results are replicated with the exception of New Zealand, where all indicators, i.e. correlations, RMSE, and MBE are better for MERRA-2. However, only the MBE shows a significant improvement when comparing MERRA-2 with ERA5. 
The differences in correlations between countries indicate that the ERA5 based simulation in most regions has a higher correlation than the one based on MERRA-2, except for New Zealand (see also Figure \ref{fig:era5_vs_merra2}). 
In summary, using ERA5 as data source for wind power simulations will result in better or at least as good timeseries as using MERRA-2. On average, quality indicators are reasonable, but extreme outliers are observed for both data sets. As they mostly occur for both reanalysis data sets, this may also be a problem of lacking data quality in observed wind power generation.

\subsection{Bias correction with GWA}
\begin{figure}[!h]
\centering
\includegraphics[scale=0.7]{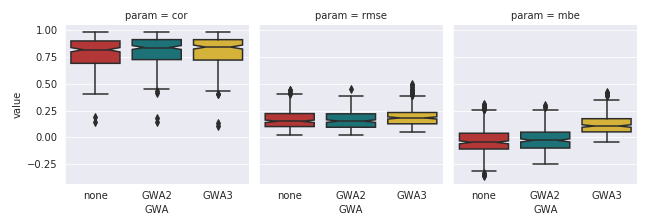}
\caption{Comparison of statistical parameters for simulations with ERA5 and different versions of the GWA for all analysed regions. Non-overlapping notches indicate difference in medians statistically significant at the 95\% significance level.}
\label{fig:era5_gwa_all}
\end{figure}

In order to adjust the mean bias of the wind speeds taken from reanalysis data, we use the Global Wind Atlas. Due to the higher spatial resolution compared to the reanalysis data sets, we expect an improvement in particular in RMSE and MBE. The effect of bias-correction on correlations depends on the non-linear relationship between wind speeds and wind power as shifting wind speeds by a constant factor does not imply a proportional shift in wind power output. Hence, bias correction may impact correlations, too. In most cases, however, this impact is small and not significant (see \ref{fig:era5_gwa_all}). In New Zealand, correlations are slightly increased with GWA2 and in South Africa using any of the GWAs, however these increases are not significant (Figure \ref{fig:era5_gwa}). \\

The RMSEs are decreased slightly by GWA2 in comparison to simulations without bias correction, but the median does not differ significantly. The simulation with GWA3, however, implies a significant increase of the median of the distribution of RMSEs, compared to GWA2 as well as compared to the simulation without mean bias correction. On a regional level, however, the significant difference in medians of GWA3 to the other simulations is only found in the USA, as well as between simulations with GWA2 and GWA3 in New Zealand (see Figure \ref{fig:era5_gwa}), i.e. the overall results are mainly driven by the US and New Zealand.\\
If measured by MBEs, a similar conclusion can be drawn: GWA2 reduces the median of the error and shifts it closer to 0. Even though this is not significant for the overall regions, a significant shift towards 0 is seen in all countries besides New Zealand.
The GWA3, in contrast, leads to a large increase in the MBE. This applies also in New Zealand and South Africa, while for Brazil GWA2 is less recommended.\\
To sum up, in most of the investigated regions, the GWA2 may be used to increase correlations (New Zealand, South Africa), decrease the RMSE (all countries) and shift the MBE closer to 0 or to a small positive value (all except Brazil). From our results, GWA3 is not recommended for bias correction as it increases the errors (RMSEs as well as MBEs for three out of four countries, see Figure \ref{fig:era5_gwa_all}).\\
A similar analysis was conducted by applying the GWA to MERRA-2 based wind power simulation. The results can be found in section \ref{subsection:merra2_gwa}. For MERRA-2, using the GWA for bias-correction has ambiguous impacts on results and we therefore do not fully recommend using it as a mean for bias-correction.\\

\subsection{Impact of spatial and temporal aggregation}
\begin{figure}[!h]
\centering
\includegraphics[scale=0.7]{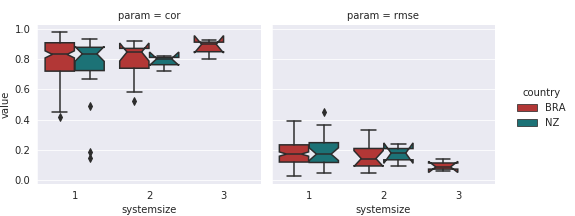}
\caption{Impact of spatial resolution (system size 1: wind parks (system size parameter (ssp) $<$ 5), system size 2: states of Brazil and New Zealand (5 $\leq$ ssp $<$ 25), system size 3: Brazil (ssp $\leq$ 25)) on simulation quality in Brazil and New Zealand. Non-overlapping notches indicate a statistical difference in the median at the 95\% significance level.}
\label{fig:spatial_res_all}
\end{figure}

In this section we assess the impact of spatial and temporal aggregation on the quality of wind power simulations. The impact on the correlation cannot be analytically derived: while an aggregation of two time-series of capacity factors will lower the variance of the combined time-series compared to the maximum of the variance of the original time-series, the change in co-variance of the combined time-series compared to the single locations cannot be analytically derived, as it depends on the co-variances of wind patterns at the two locations (see Appendix \ref{subsection:aggregation_timeseries}).\\
Therefore, we assess here empirically, how aggregation impacts time-series quality. For this analysis, the wind power simulations with ERA5 data and bias correction with GWA2 on Brazil and New Zealand (the only countries in which wind park level data are available)) are used, as this combination showed decent simulation quality for all regions.
Figure \ref{fig:spatial_res_all} shows the resulting simulation quality indicators. Overall, a tendency that at larger system size, the simulation quality as measured by correlations and RMSEs decrease can be observed. In particular, the largest system (Brazil) has a significantly lower median than the smaller systems in terms of RMSE, although single negative outliers can reach the simulation quality of the largest systems. For particular countries, this is difficult to assess, since there is a lack of variety of different system sizes. Nevertheless, in the USA and Brazil simulation quality increases as can be observed in Figure \ref{fig:spatial_res}.

With regard to spatial relations, we also assess how geography might impact the accuracy of simulation. We therefore consult the correlations of the best simulation (ERA5 with GWA2 mean bias correction) in Brazil and New Zealand (where validation data on wind park level are available). Figure \ref{fig:map_corr} indicates that in Brazil southern wind parks have higher correlation, whereas in New Zealand the highest correlations are found in proximity to the coast.

\begin{figure}[ht]
\centering
\includegraphics[scale=0.7]{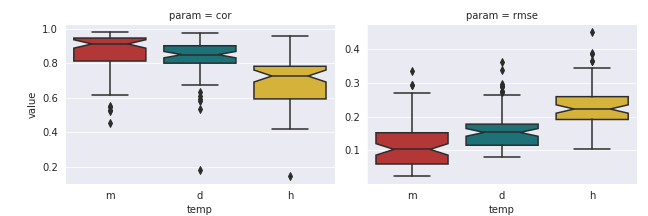}
\caption{Impact of temporal resolution on simulation quality. Non-overlapping notches indicate a statistical difference in the median at the 95\% significance level.}
\label{fig:temporal_res_all}
\end{figure}

When assessing the impact of temporal resolution on simulation quality, for the US some locations had to be excluded, as they do not provide hourly time resolution. Therefore, there only the regions of Texas and the Bonneville Power Administration were included. In all other countries, all locations are available at hourly resolution.
The medians of correlation significantly increase from hourly to daily as well as daily to monthly correlations (Figure \ref{fig:temporal_res_all}. While the increase from daily to monthly correlation is at around 5 \% points, daily correlations are around 15 \% points higher than hourly ones. This is observed in all individual countries, however only Brazil shows significant changes in median correlation for both temporal aggregation steps (Figure \ref{fig:temporal_res}).\\
The RMSE can be reduced by temporal aggregation, from hourly to daily by about 12 \% points, and from daily to monthly by around 10 \% points on average. In all countries except Brazil, the decrease in correlation is significant (Figure \ref{fig:temporal_res}).

To sum up, simulation quality tends to increase rather strongly when aggregating temporally. Spatial aggregation is somehow ambiguous, but when comparing very low to very high resolutions, the effect can also be detected.

\section{Discussion}
In this work we compare the capabilities of the two reanalyses MERRA-2 and ERA5 as data sources for wind power simulation in several countries around the world and analyse the suitability of the Global Wind Atlas to increase the quality of the simulated time series.
With a few exceptions, ERA5 performs better, with respect to the chosen quality measures and the selected samples, than MERRA-2. The better performance may be partly due to a higher spatial resolution of the input data set, but also due to using a more recent climate model based on a large amount of observed data \cite{era5_indat}. The capability of representing wind conditions especially in complex terrain should therefore be improved \cite{Olauson_2018}. 
This result is not supported by Lileó et al. \cite{lileo2013long} who claim that an increase in spatial resolution does not necessarily result in higher correlations between reanalyses and local wind measurements in a similar assessment for wind speeds.
Our results coincide with findings of Olauson \cite{Olauson_2018}, who studied the performance of these two reanalysis data sets for wind power simulation in four European countries and a region in the USA, as well as Jourdier \cite{Jourdier2020} who compared MERRA-2, ERA5, two high-resolution models and the New European Wind Atlas for the purpose of wind power simulation in France.
Olauson found hourly correlations of over 0.94 for all regions investigated (except the BPA with MERRA-2, where it is at 0.75), which is higher than the correlations identified in our study. For most locations, we find correlations above 0.7, only in South Africa they are around 0.6 (ERA5) or even below (MERRA-2). This coincides with the correlations fround by Olauson for individual wind parks in Sweden, which are above 0.5 (MERRA-2) and 0.8 (ERA5).
While Olauson finds an increase in correlation by ERA5 compared to MERRA-2 by less than 1 \% point in three of the examined regions (i.e. Germany, Denmark and France), in our study correlations of ERA5 are up to 10 \% points higher, with a higher increase in some exceptional cases. This is in the range of the increase in correlation reported by Jourdier \cite{Jourdier2020} in France and sub regions, with correlation being 0.15 higher for ERA5 compared to MERRA-2. However, in our analysis in some cases there is also a lower correlation with ERA5 based simulations compared to MERRA-2, especially in New Zealand. An interesting result is that in \cite{Olauson_2018} the highest increase in correlation by nearly 20 \% points is seen in the BPA in the USA, which agrees with the results of the present study.

Only for the USA we estimated RMSEs comparable to the results in \cite{Olauson_2018}, with values between 2.35 \% and 9.1 \% for ERA5, and 2.82 \% and 18.4 \% for MERRA-2. In the other regions (Brazil, New Zealand, South Africa), the RMSE is higher, with about 75 \% of the locations showing RMSEs above 10 \%. Reasons for these differences may be explained on the one hand by different data quality of validation data, on the other hand by a better fit of the data for the regions of the USA and Europe compared to other world regions (South America, Africa or Oceania).
Regarding the comparison of the two reanalyses, Olauson found that for ERA5, the RMSE was between 20 \% and 50 \% lower than for MERRA-2 (except in Denmark where there was hardly any impact). In absolute terms, this means a decrease of up to 0.02 (except for BPA with over 0.09), while we found that in some locations the RMSE was up to 0.2 lower for ERA5 than for MERRA-2. In other, but fewer locations, particularly in New Zealand, however, the RMSE was up to 0.2 higher with ERA5 compared to MERRA-2 based simulations.

The GWA does not improve simulation quality consistently for all locations. While GWA2 showed a potential to decrease RMSEs, GWA3 rather increases them. Considering the MBEs, the results are ambiguous. GWA3 often increased errors and performed worse than GWA2. Despite an analysis showing that ERA5 performs better than ERA-Interim \cite{Rivas2019}, this cannot be confirmed for GWA3 and GWA2, respectively, which are based on these two different reanalysis data sets. So far, no other study using the GWA3 has been conducted, but results from analyses of the previous version showed that applying the GWA for downscaling MERRA reanalysis wind speeds (EMHIRES dataset \cite{gonzalez2016emhires}) has no unambiguously positive effect on the simulation quality when compared to TSO time series. Despite the claim of the authors that the simulation based on MERRA data underestimates the variability compared to the GWA-downscaled dataset (EMHIRES) and that downscaling improves results, their statistical results indicate that neither correlations increase (13 of 24 countries investigated have higher correlation with EMHIRES than with MERRA), nor RMSE (9 countries) or biases (7 countries) decrease consistently \cite{GonzalezAparicio2017}. This fits well to the results of our current study, where the results of different countries or regions vary in terms of whether the GWA improves the quality of wind power simulation time series or not. Another study which uses the GWA and MERRA-2 for wind power simulation in Brazil finds that bias correction in general improves results \cite{Gruber2019}.

A further subject we investigated are the implications of spatial and temporal aggregation on the measures applied for quality assessment. The expectation was that the higher the level of spatial or temporal aggregation, the lower the error, since compensating effects of negative and positive bias could reduce errors. For temporal aggregation this could be confirmed by the analysed data. This is also confirmed by Staffell and Pfenninger who compute higher correlations for eight European countries on a monthly than on an hourly basis \cite{Staffell_2016} .
For spatial aggregation, however, we could not consistently confirm such an effect. This matches the results of an analysis conducted in Europe, using MERRA and MERRA-2 reanalysis data. Monthly correlations on country level were lower than correlations on European level only in some of the 13 studied countries (9 for MERRA and 7 for MERRA-2). Also, the median of correlations per country was above the correlations of aggregated data \cite{Staffell_2016}. In contrast to this Olauson \cite{Olauson_2018} finds higher correlations, as well as lower RMSEs and errors in Sweden compared to 1051 individual wind turbines when simulating wind power with MERRA-2 and ERA5.
Limitations of this study were data availability and data quality. For future research, also validation in other countries is desirable. Moreover, better quality data for simulation could highly increase the validity of the results. Nevertheless, we feel confident that our results hold when comparing different simulations, despite some of the validation timeseries being of lesser quality.\\

\section{Conclusions}
In this paper we assessed how different reanalysis data sets for wind power simulation in different regions of the world, as well as means for global bias correction of reanalysis wind speeds, affect simulation quality. We additionally looked into the implications of spatial and temporal aggregation on quality measures.

Our main conclusions are (1) that ERA5 performs better than MERRA-2 in all regions and for all different indicators, with ERA5 showing approximately 0.05 higher correlations than MERRA-2 and 0.05 lower RMSEs in most regions. (2) No version of the GWA consistently improves simulation quality. GWA2 may be used, however improvements over the use of no bias correction may be minor and in some cases, simulation results may even deteriorate. We discourage the use of GWA3. (3) Temporal aggregation increases quality indicators due to compensating effects, with an increase of about 0.2 in correlation and about 0.1 to 0.2 lower RMSEs in most regions when aggregating from hourly to monthly time series. (4) For spatial aggregation, a much more limited effect was found: only when comparing very low and very high spatial aggregations, an increase in quality was observed.\\

The results of our analysis  \footnote{The resulting time series aggregated per wind park will be made available after submission in an online repository} can be used as basis for future wind power simulation efforts and are the foundation for a new global dynamic wind atlas. Access to this global dynamic wind atlas is enabled by making our code openly available \cite{GDWA}.
The tool is able to generate wind power generation timeseries for all locations worldwide for use in energy system models or for studying the variability of wind power generation. Furthermore, our results allow estimating the magnitude of error that has to be expected when relying on reanalysis data for wind power simulation. These conclusions are important for energy system modellers when designing highly renewable energy systems. 

\section{Acknowledgements}
This project has received funding from the European Research Council (ERC) under the European Union’s Horizon 2020 research and innovation programme (grant agreement No. 758149).

\sloppy
\printbibliography

\newpage
\appendix
\section{Appendix}
\subsection{Aggregation of time series}
\label{subsection:aggregation_timeseries}
We have time series $X$ and $Y$ and measure their similarity using the correlation. We want to see if aggregation of time series has an impact on correlation.

Let be $X_{1}$ and $Y_{1}$ time series e.g. in region $1$ and $X_{2}$ and $Y_{2}$ time series in another region $2$. Given correlations $corr\left(X_{1},Y_{1}\right)$ and $corr\left(X_{2},Y_{2}\right)$ we are interested in $corr\left(X,Y\right)$ for the aggregated time
series $X:=a\cdot X_{1}+b\cdot X_{2}$ and $Y:=a\cdot Y_{1}+b\cdot Y_{2}$ for some $a,b>0$ with $a+b=1$.

Note: we are not interested in negative correlations, so when we say ``increase of correlation'' we mean that $\left|corr\left(X_{1},Y_{1}\right)\right|<\left|corr\left(X,Y\right)\right|$.

Let's first show that correlation can increase by aggregation:
\begin{example}
Let $Z$ be some arbitrary random variable with $\mathbb{V}Z\neq0$.
Further assume that $X_{1}$ and $Y_{1}$ are independent. If we set
$X_{2}:=-X_{1}+Z$, $Y_{2}:=-Y_{1}+Z$ and $a:=\frac{1}{2}$ and $b:=\frac{1}{2}$,
then we get $X=\frac{1}{2}Z$ and $Y=\frac{1}{2}Z$. Therefore $corr(X_{1},Y_{1})=0$
but $corr\left(X,Y\right)=1$. If we choose $Z$ with $\mathbb{V}Z\ll\mathbb{V}X_{1}$
and $\mathbb{V}Z\ll\mathbb{V}X_{2}$, also $corr(X_{2},Y_{2})$ is
almost $0$.
\end{example}

\begin{algorithm}
\begin{lstlisting}
import numpy as np

N = 1000
noise = np.random.normal(size=N, scale=0.1)
x1 = np.random.normal(size=N)
x2 = -x1 + noise
y1 = np.random.normal(size=N)
y2 = -y1 + noise
a = b = 0.5

def corr(x, y):
    return np.corrcoef(x, y)[0, 1]

print("x1 and y1 are not correlated:", corr(x1, y1))
print("x2 and y2 are not correlated:", corr(x1, y1))
print("x and y are strongly correlated:",
	corr(a * x1 + b * x2, a * y1 + b * y2))

# weirdly this is violated:
# min(var(x1), var(x2)) <= var(ax1+bx2) <= max(var(x1), var(x2))

print("var(x1): ", np.var(x1))
print("var(x2): ", np.var(x2))
print("var(a*x1 + b*x2): ", np.var(a * x1 + b * x2))
\end{lstlisting}

Output:

\begin{lstlisting}
x1 and y1 are not correlated: -0.003968605464354068
x2 and y2 are not correlated: -0.003968605464354068
x and y are strongly correlated: 1.0
var(x1):  1.0845349544321836
var(x2):  1.0952631334805492
var(a*x1 + b*x2):  0.00228463494772665
\end{lstlisting}

\caption{Numerical example for Example 1.}
\end{algorithm}

Now we show that correlation of the aggregated random variables can
vanish even for high correlation of $X_{i}$ amd $Y_{i}$, $i=1,2$.
\begin{example}
Now choose $Z$ to be a random variable with $0<\mathbb{V}Z\ll\mathbb{V}X_{i}$
for $i=1,2$. Further let $X_{1}$ be some arbitrary random variable
with $\mathbb{V}X_{1}\neq0$ and independent to $Z$.

Then set $X_{2}:=-X_{1}+Z$, $Y_{1}:=3\cdot X_{1}$ and $Y_{2}:=-X_{1}$.
This yields $X=\frac{1}{2}X_{1}-\frac{1}{2}X_{1}+\frac{1}{2}Z$ and
$Y=X_{1}$. Since $X_{1}$ and $Z$ was chosen to be independent,
we have $corr\left(X,Y\right)=0$, but $corr\left(X_{1},Y_{1}\right)=1$
and $corr\left(X_{2},Y_{2}\right)$ is very close to $1$ because
$Z$ was chosen to be small noise.
\end{example}

\subsection{Validation of USWTDB with IRENA}
\label{subsection:uswtdb_irena}
\begin{figure}[ht]
\centering
\includegraphics[scale=0.7]{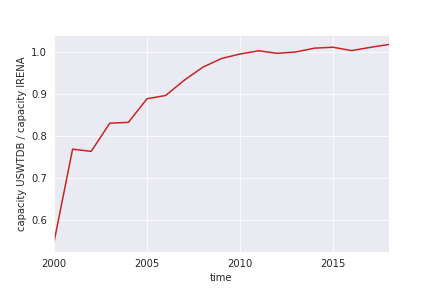}
\caption{Installed capacities in the US wind turbine data base \cite{USWTDB} compared to IRENA \cite{IRENA}}
\label{fig:uswtdb_irena}
\end{figure}

We validate the data in the USWTDB \cite{USWTDB} with installed capacities as provided by the International Renewable Energy Agency (IRENA) \cite{IRENA}. Figure \ref{fig:uswtdb_irena} shows the ratio of capacities in the USWTDB to IRENA capacities. After the year 2010, this ratio is close to 1, but before 2010 capacities do differ quite significantly. This indicates that there are large capacities missing in the USWTDB in earlier years.

\subsection{Additional data sources South Africa wind parks}
For the wind park data set in South Africa, part of the information was missing and therefore needed to be complemented by additional sources. Some data points were not available at all and data was selected according to turbine specific available data. The turbine type is known, therefore from turbine specification data sheets missing information can be derived. In case there are several possibilities, e.g. hub height in a range, instead of only one number, a value in the medium range is picked. The data that needed to be added are listed in Table \ref{tab:zaf_turb_complete}.

\begin{table}[h!]
\caption{Additional data gathered for complementing the South African wind park data set}
\makebox[\textwidth][c]{}
\begin{tabular}{p{2.5cm}p{2.8cm}p{4cm}p{6cm}}
\hline
Windpark                      & Data quality issue                                 & Correction                                                                  & Source                                                                                                         \\ \hline
Dorper                        & wrong height                          & Set to 80 m  (derived from existing turbine type)                                                            &   
Nordex N100/2500 \cite{NordexN100}\\
Excelsior                     & missing height                        & 
Set to 100 m  (derived from existing turbine type)                                                                 &Goldwind GW121/2500 \cite{GoldwindGW121}
\\
Gibson Bay                    & missing height                        & Set to 100 m (derived from existing turbine type)                                                                &
Nordex N117/3000 \cite{NordexN117}\\
Longyuan                      & once height, once diameter missing    & complement with each other                                               &                                                                                                                \\
Karusa                        & missing height and diameter           & use height from project homepage                                         & https://www.windbase.eu/projects/wind-farm-karusa-en-soetwater.aspx                                            \\
Soetwater                     & missing height                        & use height from project homepage                                         & https://www.windbase.eu/projects/wind-farm-karusa-en-soetwater.aspx                                            \\
Tsitsikamma                   & missing height                        & set to 112 m   (derived from existing turbine type)                                                              &    ref                                                                                                            \\
Wesley-Ciskey                 & missing height, diameter and capacity & assume 126 m diameter and 137 m height for V126 3.45MW                   & https://www.afrik21.africa/en/south-africa-vestas-to-build-wesley-ciskei-wind-farm-for-edf/                    \\
Nxuba                         & missing height, diameter and capacity & assume Acciona AW123-3MW with 125 m diameter and 120 m height            & https://www.aced.co.za/nxuba-wind-farm                                                                         \\
Oyster Bay                    & missing height and diameter           & assume Vestas V117-3.45 with 117 m diameter and 91.5 m height            & https://www.aa.com.tr/en/energy/news-from-companies/vestas-awarded-148-mw-wind-project-in-south-africa/22153   \\
Klawer Wind Farm              & missing height, diameter and capacity & assume information from project plan                                     & http://www.energy.gov.za/files/esources/kyoto /2011/06-06-2011\%20-\%20Klawer\%20PDD.pdf                        \\
Hopefield Community Wind Farm & missing height, diameter and capacity & assume same as Hopefield Wind Farm                                       &                                                                                                                \\
Golden Valley                 & missing height, diameter and capacity & assume GW121/2500 with 121 m diameter, 120 m height and 2500 kW capacity & https://www.windpowermonthly.com/article /1488858/long-delayed-south-african-wind-farms-reach-financial-close   \\
Garob                         & missing height and diameter           & assume AW125/3150 with 125 m diameter and 120 m height                   & https://www.afrik21.africa/en/south-africa-enel-begins-garob-wind-farm-construction-140-mw/                    \\
Copperton                     & missing height and diameter           & assume AW125/3150 with 125 m diameter and 120 m height                   & https://www.evwind.es/2018/09/13/nordex-acciona-awarded-big-ticket-wind-energy-contracts-in-south-africa/64501 \\ \hline
\end{tabular}
\label{tab:zaf_turb_complete}
\end{table}

\subsection{Difference in simulation quality MERRA-2 vs. ERA5}
\begin{figure}[ht]
\centering
\includegraphics[width=\linewidth]{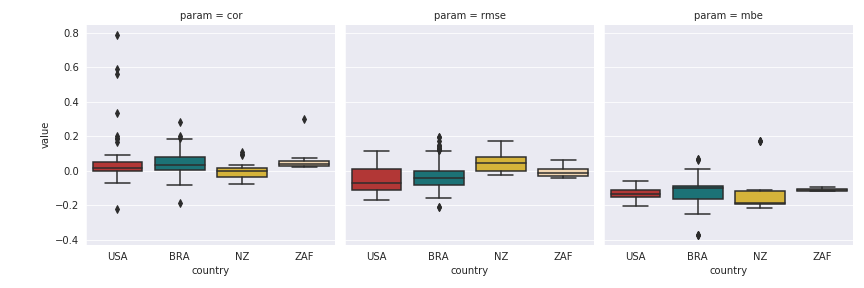}
\caption {Differences (ERA5 - MERRA-2) in statistical parameters for simulations with MERRA-2 and ERA5 (MERRA-2 - ERA5)}
\label{fig:era5_vs_merra2_dif}
\end{figure}

Figure \ref{fig:era5_vs_merra2_dif} displays the change in the indicators correlation, RMSE and MBE when applying ERA5 instead of MERRA-2 for wind power simulation. In the USA, Brazil and New Zealand the correlation is up to 10 \% points higher with ERA5 than with MERRA-2, for the USA there are even some outliers with an increase in correlation of up to 80 \% points. Only in New Zealand correlations with ERA5 based simulations are lower.
The RMSE is lower with ERA5 compared with MERRA-2 except in New Zealand where simulations with ERA5 result in RMSEs up to 20 \% points higher than with MERRA-2. The difference in MBEs is more consistent in the different regions, in the range of 10 to 20 \% points lower for ERA5.

\subsection{Applying GWA to MERRA-2 simulated wind power time series}
\label{subsection:merra2_gwa}
\begin{figure}[ht]
\centering
\includegraphics[width=\linewidth]{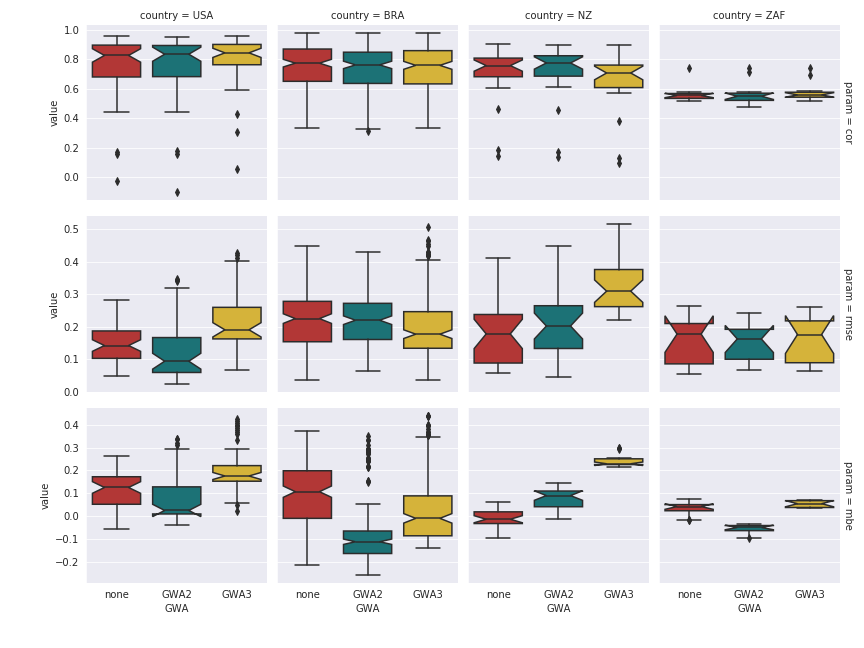}
\caption {Comparison of statistical parameters for simulations with MERRA-2 and different versions of the GWA}
\label{fig:merra2_gwa}
\end{figure}

Here, we show the impacts of applying GWA to MERRA-2 data. As for ERA5, in most cases the impact of the GWA on correlations is negligible, as can be seen in Figure \ref{fig:merra2_gwa}. In New Zealand the correlation is slightly increased with GWA2 and decreased with GWA3, but the changes are not significant. The RMSE decreases with GWA2 in all regions but New Zealand (the decrease is only significant in the USA), while GWA3 shows a tendency to increase the RMSE (with significantly increased RMSE in the USA and New Zealand) except in Brazil where it has a significantly decreasing effect.
In Brazil the best fit according to MBEs is observed using GWA 3 which decreases the MBE leading to a lower error. As with ERA5, using GWA2 decreases MBEs leading to an underestimation on average. In the USA, the smallest mean bias is achieved with GWA2 which reduces the MBE, while GWA3 increases the MBE and thus the error. In New Zealand, using no bias correction with GWA leads to a small error and a good fit. If GWA2 is applied, overestimation of around 10 \% capacity factor is achieved, while GWA3 increases the overestimation to more than 20 \% capacity factor. For New Zealand it is therefore not recommended to apply GWA for mean bias correction.
In South Africa simulations overestimate observed power generation by around 5 \% capacity factor, which is increased slightly but insignificantly by GWA3, while GWA2 decreases the error to nearly -10 \% capacity factor. The best fit is therefore achieved without GWA. All other changes in MBE are significant.\\
To sum up, the results of mean bias correction with GWA using MERRA-2 reanalysis data is ambiguous. While the RMSE is decreased except for New Zealand with GWA2, GWA3 usually increases the RMSE, but on the other hand performs better than GWA2 in terms of MBE in Brazil and South Africa. From these results, neither GWA2 nor GWA3 can be fully recommended for bias correction of MERRA-2 data as simulation quality is not consistently increased.

\subsection{Country-specific results}
This section presents results as in section \ref{section:results}, but per country. This allows to compare specifics in different countries compared to the overall picture.

\subsubsection{Impact of the choice of reanalysis dataset on simulation quality}
\begin{figure}[!h]
\centering
\includegraphics[width=\linewidth]{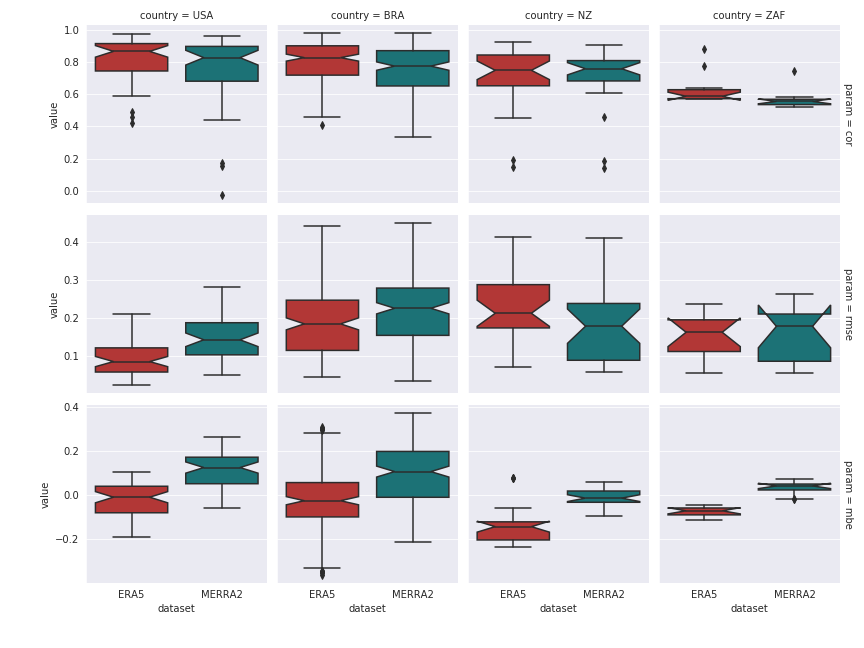}
\caption{Comparison of statistical parameters for simulations with ERA5 and MERRA-2 reanalyses for each of the four countries analysed individually. Non-overlapping notches indicate a  difference in the medians statistically significant at the 95\% level.}
\label{fig:era5_vs_merra2}
\end{figure}

Figure \ref{fig:era5_vs_merra2} shows the three indicators measuring simulation quality (i.e. correlation, RMSE and MBE) in the four different countries for the two reanalysis datasets. ERA5 has, on average, higher correlations than MERRA-2. The median differs, however, only for South Africa significantly. For the RMSE, ERA5 is significantly better than MERRA-2 in the USA and Brazil. In New Zealand and South Africa, however, no significant difference in the median of RMSEs is found. The MBEs are closer to 0 in the USA and Brazil with ERA5, however MERRA-2 performs better in New Zealand. In South Africa the MBEs indicate a similar error for both data sets, but ERA5 underestimates while MERRA-2 overestimates. All differences in the MBE are significant.\\
Overall, it can be concluded that ERA5 performs better than MERRA-2 in terms of higher correlations but lower errors, with the exception of New Zealand (Figure \ref{fig:era5_vs_merra2}). However, in many cases the differences in the median between the two datasets are insignificant (95\% confidence interval), in particular for the correlations.

\subsubsection{Bias correction with GWA}
Regarding correlations, the changes are minor, only in New Zealand there is a shift up to higher correlations with GWA2, in South Africa with both GWAs, but in none of these regions significantly.\\
The RMSE decreases with GWA2 in all regions, while the GWA3 shows a tendency to increase the RMSE. Only in Brazil the impact of the GWA3 is minor. While no version of the GWA increases or decreases the RMSE significantly, in the USA and New Zealand the simulation with GWA3 has a significantly higher RMSE than with GWA2. In the USA, the GWA2 however reduces the spread of RMSEs from between approximately 0.05 and 0.15 (IQR: 0.05) to 0.04 and 0.21 (IQR: 0.1) without GWA.\\
Regarding the MBEs, in Brazil the best fit is observed without using bias correction. With GWA2, MBEs are decreased, indicating an underestimation, while GWA3 results in an increase of MBEs. As no downtime, wake effect or other losses are taken into account in the wind power simulation model, an overestimation as with GWA3 seems more appropriate. In the USA, using no bias correction at all results in the best fit to observed wind power generation as measured by the MBE. GWA2 slightly increases the error, and GWA3 does so even more. In this case, GWA2 might be used to shift the MBE more to a positive range, to take account of possible losses. 
In New Zealand, observed wind power generation is underestimated by around 10 to 20 \% of the capacity factor without bias correction. If GWA2 is applied, generation is overestimated by up to 7 \%, while GWA3 increases the overestimation to around 15 to 20 \%. For New Zealand it is therefore also recommendable to apply bias correction with the GWA2.\\
In South Africa, simulations underestimate observed power generation by circa 10 \% capacity factor, which is decreased to less than 5 \% by GWA2, while GWA3 increases the error to nearly 10 \% capacity factor. In all studied regions, the median of MBEs differ significantly. Furthermore, in all regions the spread of MBEs is decreased when using bias correction, with the interquartile range (IQR) reducing by about 50\% except in Brazil.\\

\begin{figure}[!h]
\centering
\includegraphics[width=\linewidth]{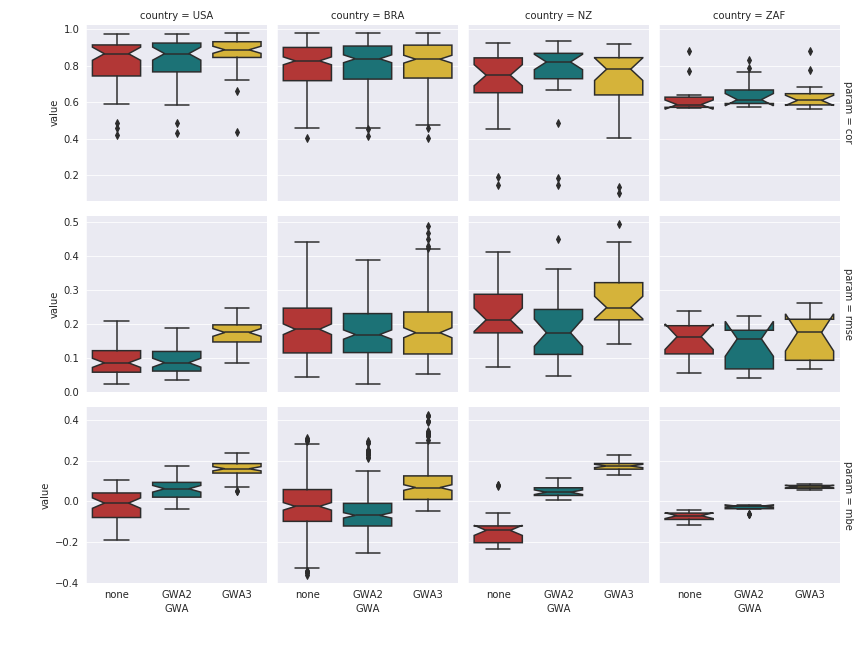}
\caption{Comparison of statistical parameters for simulations with ERA5 and different versions of the GWA. Non-overlapping notches indicate difference in medians statistically significant at the 95\% significance level.}
\label{fig:era5_gwa}
\end{figure}

\subsubsection{Wind speed correction factors}
\begin{figure}[!h]
\centering
\includegraphics[width=\linewidth]{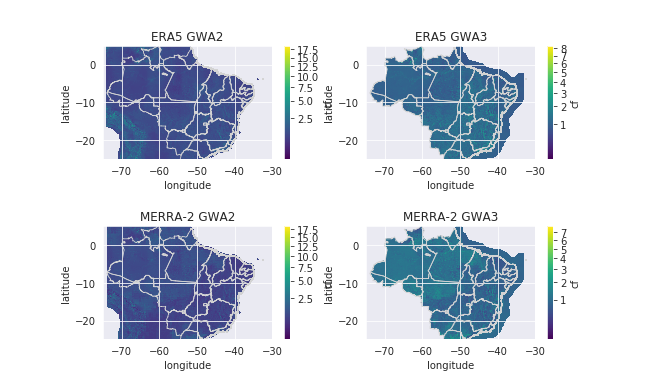}
\caption{Correction factors with GWA2 and GWA3 for MERRA-2 and ERA5 reanalyses in Brazil (the map is powerlaw-normalised)}
\label{fig:cf_BRA}
\end{figure}

\begin{figure}[!h]
\centering
\includegraphics[width=\linewidth]{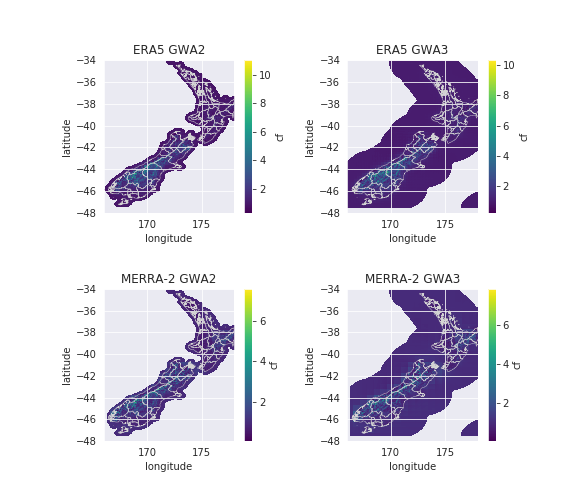}
\caption{Correction factors with GWA2 and GWA3 for MERRA-2 and ERA5 reanalyses in New Zealand}
\label{fig:cf_NZ}
\end{figure}

\begin{figure}[!h]
\centering
\includegraphics[width=\linewidth]{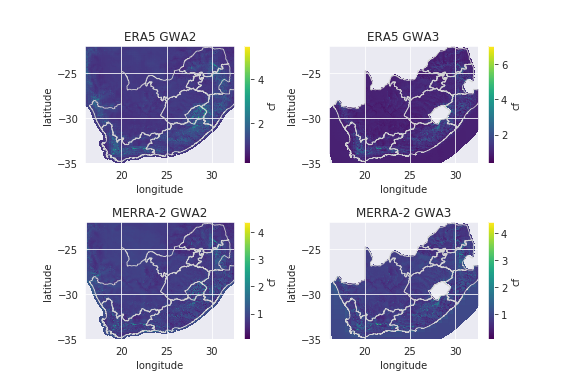}
\caption{Correction factors with GWA2 and GWA3 for MERRA-2 and ERA5 reanalyses in South Africa}
\label{fig:cf_ZAF}
\end{figure}

\begin{figure}[!h]
\centering
\includegraphics[width=\linewidth]{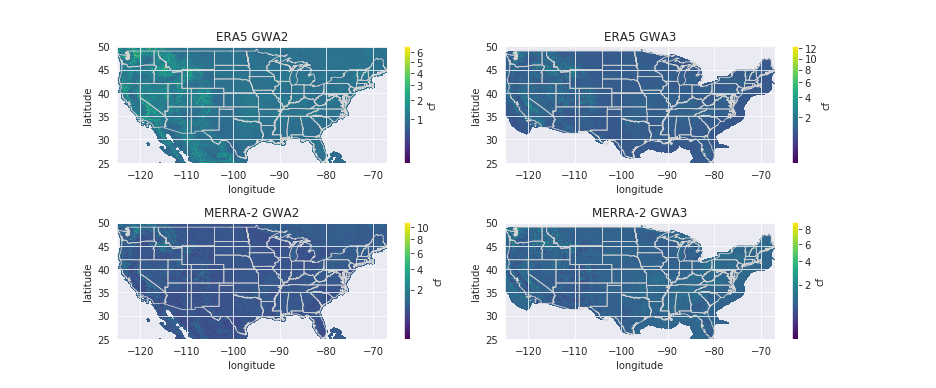}
\caption{Correction factors with GWA2 and GWA3 for MERRA-2 and ERA5 reanalyses in USA (the map is powerlaw-normalised)}
\label{fig:cf_USA}
\end{figure}

Figures \ref{fig:cf_BRA}-\ref{fig:cf_USA} show the calculated correction factors for Brazil, New Zealand, the USA and South Africa for different combinations of reanalysis and GWA datasets. A common pattern in all countries and for all datasets is that correction factors are higher in mountainous regions. Regarding the applied datasets, however, there are differences. While in New Zealand the highest correction factors are resulting form bias correction of ERA5 with any of the GWA, in the USA and South Africa this is only the case with GWA3. In the USA the correction factors with GWA2 applied to ERA5 are only about half compared to the correction factors with GWA3. In Brazil, on the other hand, the correction factors are highest with GWA2, irrespective of the reanalysis dataset they are applied on. This indicates, that either reanalysis data, or GWA, or both indicate different wind patterns depending on the region they are applied to.

\subsubsection{Relation of geography and correlations of simulated  and observed wind power generation time series}
\label{subsection:map_corr}
\begin{figure}[!h]
\centering
\includegraphics[width=\linewidth]{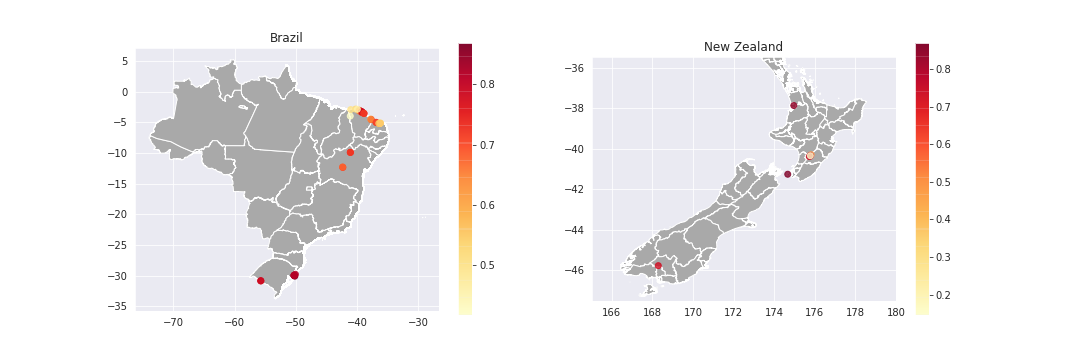}
\caption{Correlations between simulated wind power generation based on ERA5 reanalysis with GWA2 bias correction with observed wind power generation in Brazilian and New Zealand wind parks}
\label{fig:map_corr}
\end{figure}

Figure \ref{fig:map_corr} shows the hourly correlations between ERA5 simulation with GWA2 bias correction and observed wind power generation time series. In Brazil higher correlations are observed in the South, and lower at the coast of the North-East. The lowest correlations are in the north west of Ceará. In New Zealand a difference is seen between coastal and inland wind parks: At the coast correlations are higher.

\subsubsection{Impact of spatial and temporal aggregation}
For the USA, a slight tendency of higher correlations as well as lower errors (lower RMSEs, MBEs closer to 0) can be observed, when system size is increased. However, only for larger system sizes, medians of the distributions differ significantly. In Brazil, a similar trend is visible, except for the third-largest group (10, 20], in which simulation quality drops. This can be attributed to the state of Bahia, where the GWA skews the wind speeds and therefore leads to a higher over-estimation. In New Zealand and South Africa results are ambiguous, i.e. no relation between system size and simulation quality can be identified.\\

\begin{figure}[!h]
\centering
\includegraphics[width=\linewidth]{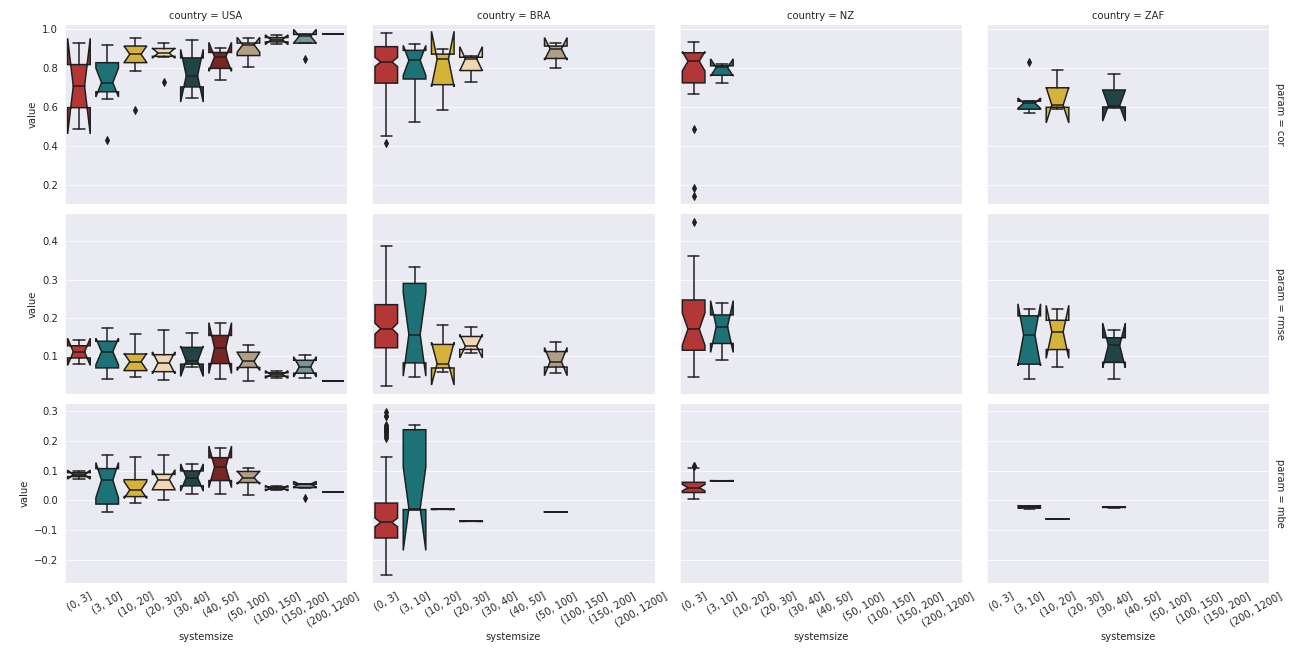}
\caption{Impact of spatial resolution (absolute system size, i.e. number of occupied reanalysis grid cells) on simulation quality per country. Non-overlapping notches indicate a statistical difference in the median at the 95\% significance level.}
\label{fig:spatial_res}
\end{figure}

\begin{figure}[ht]
\centering
\includegraphics[width=\linewidth]{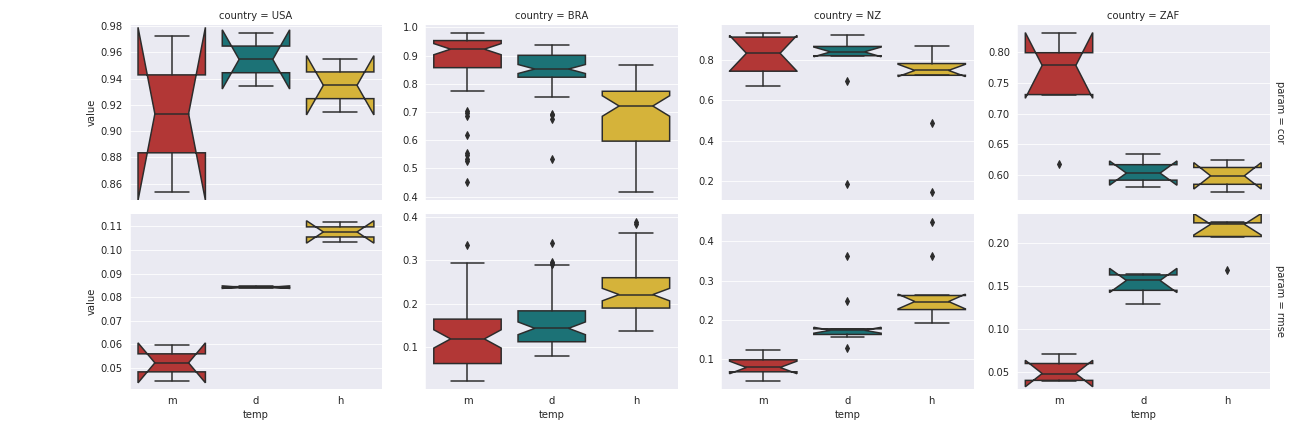}
\caption{Impact of temporal resolution on simulation quality per country. Non-overlapping notches indicate a statistical difference in the median at the 95\% significance level.}
\label{fig:temporal_res}
\end{figure}

Figure \ref{fig:temporal_res} shows a strong correlation between temporal resolution and simulation quality: as expected, the error is decreased and the correlations are increased going from hourly to monthly temporal resolution. An exception is the USA, where the monthly correlations are not higher than daily or hourly correlations. This may be a result of correlations being high for any temporal resolution in the USA ($>$ 0.85). Also, the RMSEs are the lowest (0.05 monthly to 0.11 hourly) compared to the other countries. Lowest average correlations are observed in South Africa, with hourly and daily correlations of around 0.6, which is increased to 0.75 to 0.85 by monthly aggregation. In New Zealand two very low outliers in the hourly and daily correlations are visible, which are located at The Wind Farm 3. Only in Brazil the increase by temporal aggregation  in the median of the distribution of correlations is significant, while Brazil is the only region where the RMSE does not change significantly due to temporal aggregation. The MBE is not consulted, since it is the same on average for each of the levels of temporal resolution (Figures \ref{fig:spatial_res} and \ref{fig:temporal_res}).\\

\subsection{Time series quality assessment for USA}
\label{subsection:quality_USA}
As described in section \ref{subsection:data_cleaning}, generation data in the USA were selected by visual assessment of the time series. For this purpose, the monthly time series were plotted and screened for obvious errors, such as several months of (nearly) constant wind power generation, or observed generation fluctuating between a limited amount of levels without showing typical seasonal pattern.\\

\begin{figure}[ht]
\centering
\includegraphics[width=\linewidth]{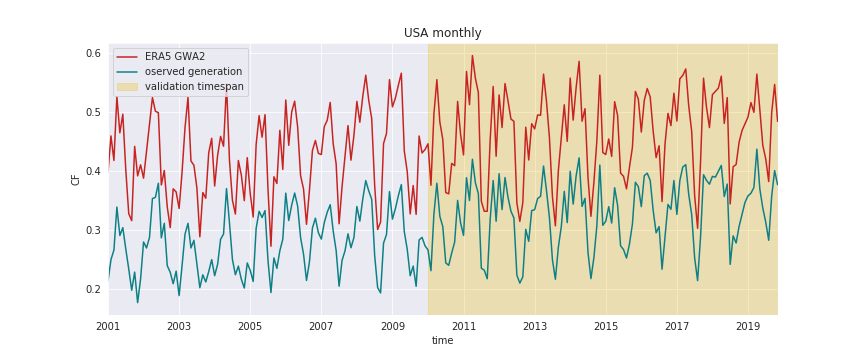}
\caption{Simulated and observed monthly wind power generation in the USA}
\label{fig:USAm}
\end{figure}

\begin{figure}[ht]
\centering
\includegraphics[width=\linewidth]{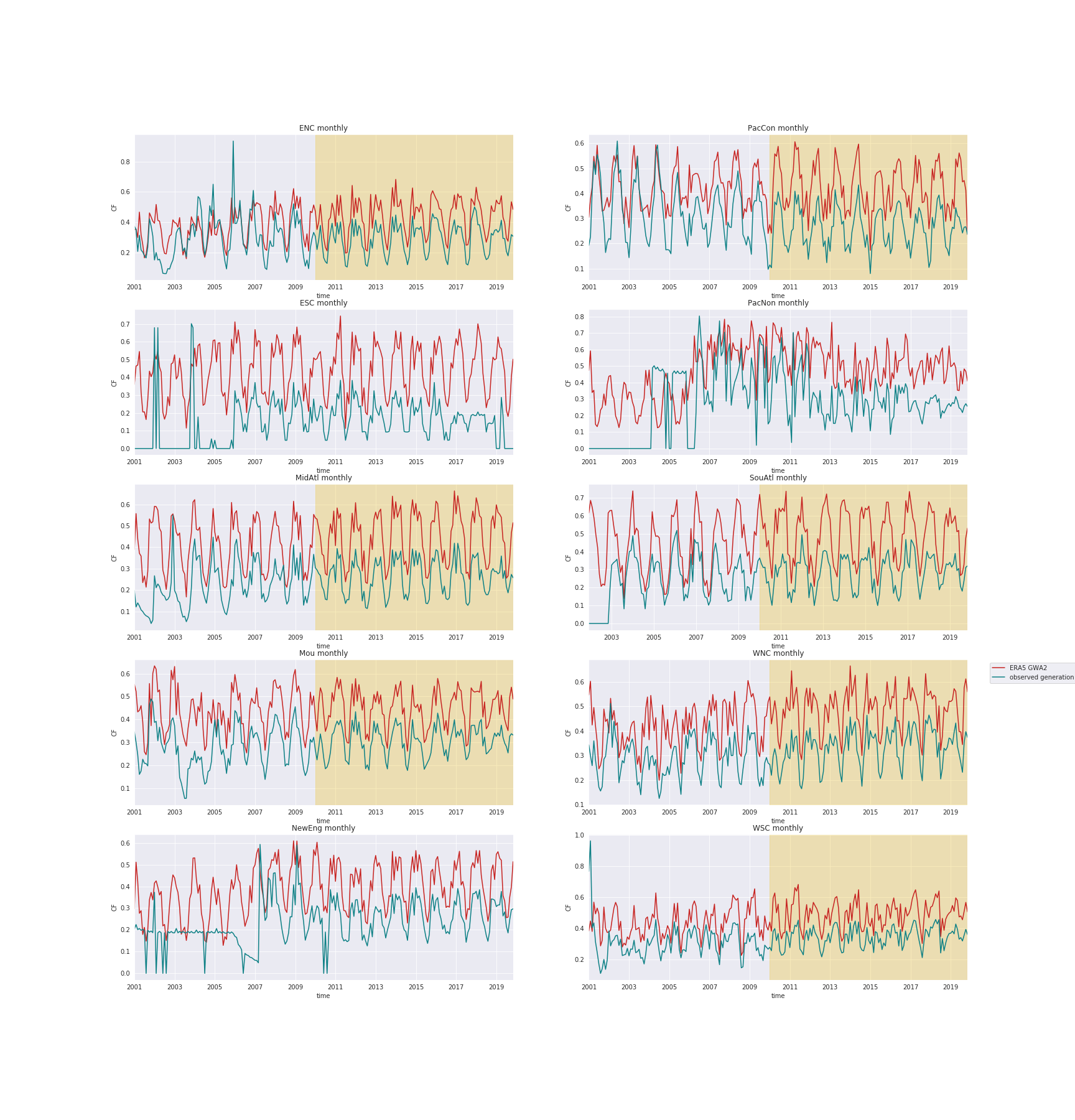}
\caption{Simulated and observed monthly wind power generation in ten regions of the USA}
\label{fig:regionsm}
\end{figure}

While the monthly generation of the USA (Figure \ref{fig:USAm}) exhibits no obvious data quality issues, three regions are removed, since their production is constant at  two nearly constant levels for the first years (Figure \ref{fig:regionsm}): New England (NewEng), East South Central (ESC) and Pacific Non-Continental (PacNon) regions.\\
Seven of the states were discarded for further use, due to their unsatisfying data quality (Figure \ref{fig:statesm}): Arkansas (AK), Connecticut (CT), Delaware (DE), Illinois (IL), North Carolina (NC), South Dakota (SD) and Tennessee (TN). In nine states only a part of the time series was used, while the the remainder was discarded due to unusual patterns such as fluctuating generation between plateaus or unusually high or low generation instead of seasonal patterns: Massachusetts (MA),  Nebraska (NE), New Jersey (NJ), Ohio (OH), Rhode Island (RI), Vermont (VT) and Wisconsin (WI).

\begin{figure}[h]
  \centering
  \setkeys{Gin}{width=\linewidth}
  \subfloat[]{\includegraphics{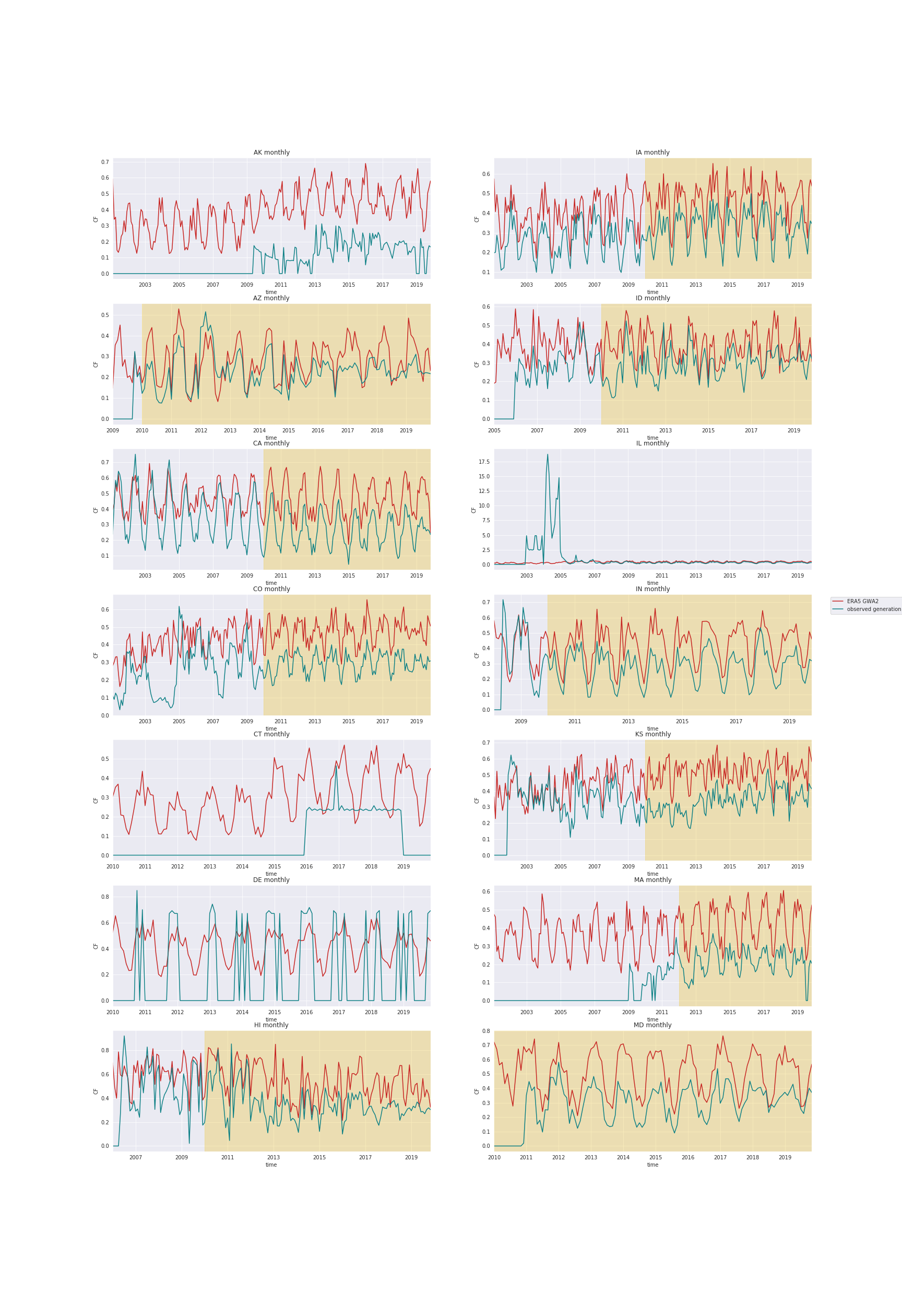}}
  \caption{Simulated and observed monthly wind power generation in the states of USA. The validation period is shaded yellow. If the time-series was not used at all, this period has 0 length.}
  \label{fig:statesm}
\end{figure}

\begin{figure}[h]
  \ContinuedFloat\centering
  \setkeys{Gin}{width=\linewidth}
  \subfloat[]{\includegraphics{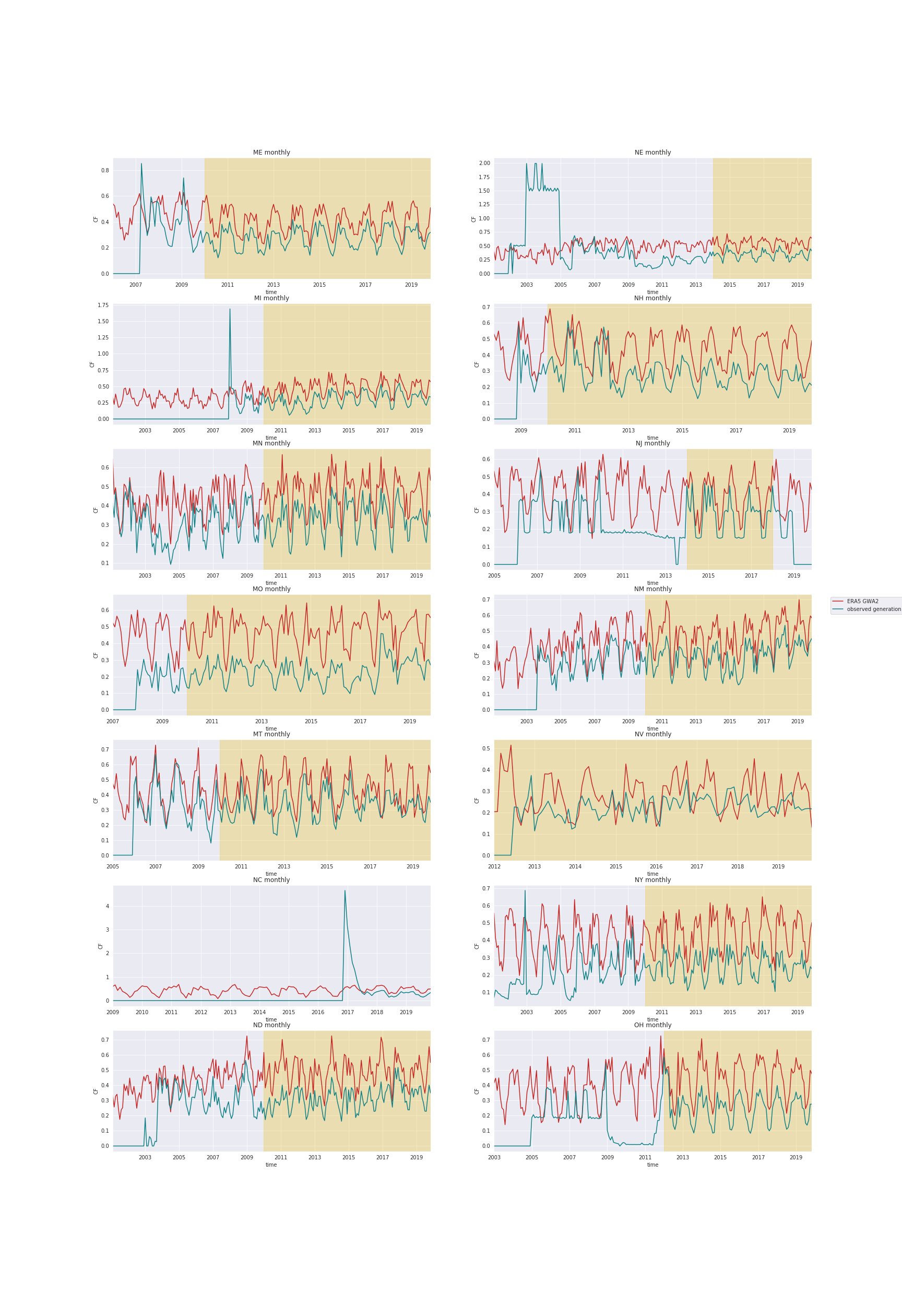}}
  \caption{Simulated and observed monthly wind power generation in the states of USA. In states where only part of the time series is used, the validation period is shaded yellow}
\end{figure}

\begin{figure}[h]
  \ContinuedFloat\centering
  \setkeys{Gin}{width=\linewidth}
  \subfloat[]{\includegraphics{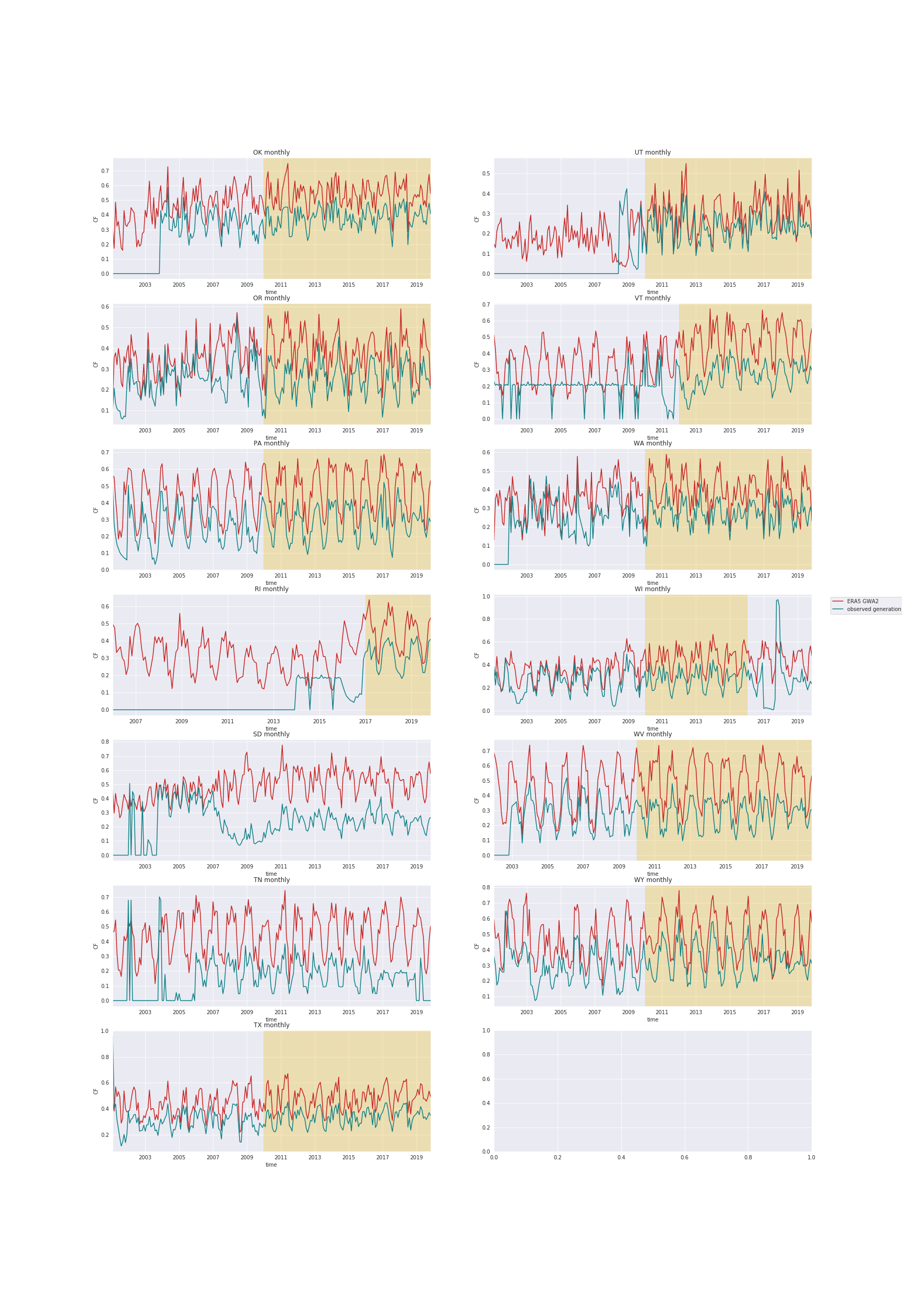}}
  \caption{Simulated and observed monthly wind power generation in the states of USA. In states where only part of the time series is used, the validation period is shaded yellow}
\end{figure}

\begin{figure}[ht]
\centering
\includegraphics[width=\linewidth]{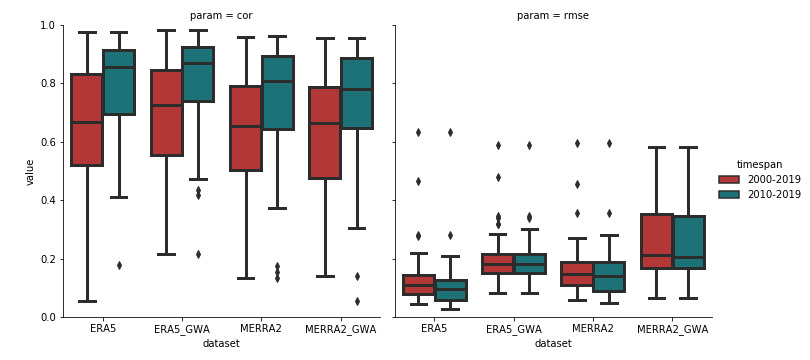}
\caption{Correlations (left) and RMSEs (right) of simulated vs. observed wind power generation in the USA and its states and regions, comparing time series for the entire period (2000-2019) to only the past ten years (2010-2019)}
\label{fig:2000vs2010}
\end{figure}

Apart from several regions with bad quality time series, it was also perceived that the observations in the years before 2010 fit the simulations worse than the past ten years. Therefore, the time series before 2010 were discarded and the results compared to the analysis based on the entire time series. As Figure \ref{fig:2000vs2010} shows, correlations can be increased and RMSEs decreased when considering the shorter period only.

\end{document}